\documentclass[%
 ,twocolumn%
,amssymb, amsmath,nobibnotes, aps, prl]{revtex4-1}
\usepackage{bm}%
\expandafter\ifx\csname package@font\endcsname\relax\else
 \expandafter\expandafter
 \expandafter\usepackage
 \expandafter\expandafter
 \expandafter{\csname package@font\endcsname}%
\fi

\usepackage{graphicx,subfigure}
\usepackage{amssymb,amsmath,wasysym}

\DeclareGraphicsExtensions{.pdf}
\graphicspath{{./figs/}}
\usepackage[usenames,dvipsnames]{color}   

\usepackage[normalem]{ulem} 

\def\be{\begin{equation}}
\def\ee{\end{equation}}
\def\bpm{\begin{pmatrix}}
\def\epm{\end{pmatrix}}

\begin{document}

\title{ Colliding cells:  when active segments behave as  active particles} 
\author{ P. Recho$^{1,*}$, T. Putelat$^{2}$  and L. Truskinovsky$^3$}
\affiliation{
$^1$ LIPhy, CNRS--UMR 5588, Universit\'e Grenoble Alpes, F-38000 Grenoble, France\\
$^2$DEM, Queen's School of Engineering, University of Bristol, 
Bristol, BS8 1TR, United Kingdom\\
$^3$PMMH, CNRS--UMR 7636, ESPCI ParisTech, F-75005 Paris, France
}

\date{\today}%

\begin{abstract} \small 
Quantifying the outcomes of cells collisions is a crucial step in building the foundations of a kinetic theory of living matter. Here, we develop a mechanical theory of such collisions by first representing individual cells as extended objects with internal activity and then reducing this description to  a model of size-less active particles characterized  by their position and polarity. We show that, in the presence of an applied force, a cell can either be dragged along or self-propel against the force, depending on the polarity of the cell. The co-existence of these regimes offers a self-consistent mechanical explanation for cell re-polarization upon contact. We rationalize the experimentally observed collision scenarios within the extended and particle models and link the various outcomes with measurable biological parameters. 
\end{abstract}

\maketitle

Interacting  cells is an important example of  active matter and the modeling of  their collective behavior  is  one of the main  challenges for non-equilibrium statistical mechanics \cite{marchetti2013hydrodynamics}. Since the  configuration of emerging  active phases  crucially  depends on  cell-to-cell collisions \cite{camley2017physical, hakim2017collective}, the rationalization of collision tests is a prerequisite for the development of an adequate kinetic theory of living matter \cite{Vicsek1995novel, peshkov2014boltzmann}. Building reliable links between the collison outcomes and  measurable biophysical parameters is also fundamental for the control of development, integrity and regeneration of living organisms \cite{FriDar_nrm09,Trep_tcb11,Gov_nmat11,Tre_nphys09,Gov_hfspj09,Kab_jrsi12,Tam_nmat11, garcia2015physics,lober2015collisions,zimmermann2016contact,smeets2016emergent}. 

Experiments show that head-on collision of two polarized cells can result in four possible outcomes~\cite{desai2013contact, scarpa2013novel}: velocity reversal, representing a quasi-elastic collision with  symmetric re-polarization,  two  quasi-inelastic scenarios with the formation of  a cell doublet  that can be either  motile (train) or static (still) and finally, a bypass regime, when cells advance over each other \cite{kulawiak2016modeling}.  

The reversal and pairing regimes are usually associated with the phenomenon of contact inhibition of locomotion  (CIL) \cite{stramer2017mechanisms}.  While CIL is crucial for healthy animal physiology, being a critical driver of cell dispersion, tiling and collective motility within embryonic tissues, the loss of CIL is usually associated with  pathological processes,  including  cancer \cite{stramer2017mechanisms}. The physical conditions provoking the failure of  CIL remain largely unknown. 

Many important advances have been made  in the  modeling of individual cells migration \cite{JulKruProJoa_pr07,rubinstein2009actin,shao2010computational,ziebert2011model, giomi2014spontaneous,tjhung2015minimal}. The extra-complexity of cell collision is due to the involvement of additional mechanisms including coordinated bonding and re-polarization. The biological control of these processes may  be complex, for instance, re-polarization has been recently modeled in reaction-diffusion frameworks with the focus on Rho-GTPase \cite{rappel2017mechanisms}. This and other bio-chemical regulators of force production inside the cytoskeleton have been already incorporated in  computational models of cell collision \cite{merchant2017rho,kulawiak2016modeling}. 
  
Motivated by the experimental observations that a crucial building block of CIL  is mysoin contractility \cite{davis2015inter}, we  take an alternative path and  study the possibility of capturing the known outcomes of a collision within a purely mechanical model. We  first represent cells as extended segments of active gel (AS)~\cite{RecPutTru_prl13,CalJonVoi_njp13} and then reduce this model to obtain an equivalent active particle (AP) description. In contrast to some well known representations of size-less active agents~\cite{romanczuk2012active},  the derived particle model contains an internal variable describing cell polarity which can be affected by the applied force. We show that this reduced AP model is able to adequately reproduce the outcomes of collision tests predicted by the AS model covering the whole set of possibilities observed experimentally.

Being exposed to an external force, both models support two coexisting dynamic regimes: \emph{frictional}, when the active object is dragged by the force, and \emph{anti-frictional}, when it is dragging  the force.  The fact that the  system can  jump from one of these nonequilibrium steady states to the other through a hysteresis loop offers a  self-consistent  mechanical explanation for  cell re-polarization upon contact.  In this description, re-polarization emerges as a result of the spontaneous self-organization of the cytoskeleton rather than an outcome of  chemical regulation.  The most important  prediction of the  model is that all four known cell collision scenarios  can be accessed by tuning a single nondimensional parameter describing cell contractility.  

We begin by representing a self-propelling cell as an AS which allows us to  focus  on the dynamics of its cytoskeletal meshwork. We use non-dimensional variables and assume that the segment is limited   by  two time-dependent fronts $x_r(t)<x_f(t)$ and has  a fixed length $\mathcal{L}=x_f-x_r$.  In this simplified setting, both fronts are moving at the same velocity $V(t)=\dot S$ where  $S(t)=(x_f+x_r)/2$ is the center of the  segment and the superimposed dot denotes the time derivative. We further assume that the organization of the molecular motors  in the segment is governed by the  dimensionless drift-diffusion  equation
\begin{equation}\label{eq:model_motors}
\dot{c}+\partial_y\left(cv\right)=\partial_{yy}c,
\end{equation} 
where  $y \in [-\mathcal{L}/2,\mathcal{L}/2]$ is a  spatial coordinate labeling material points of the moving segment, $c(y,t)$ represents the concentration of motors and $v(y,t)$ is the mechanical velocity  of the cytoskeleton relative to the motion of the cell fronts. We supplement \eqref{eq:model_motors} with no-flux boundary conditions $\partial_yc\vert_{\pm \mathcal{L}/2}=0$ insuring that the total amount of motors is conserved: $ \langle c \rangle= 1$, where spatial averaging is denoted $\langle h \rangle=\mathcal{L}^{-1}\int_{-\mathcal{L}/2}^{\mathcal{L}/2}h(z,t)dz$.

\begin{figure}[h!]
\centering
\includegraphics[scale=0.35]{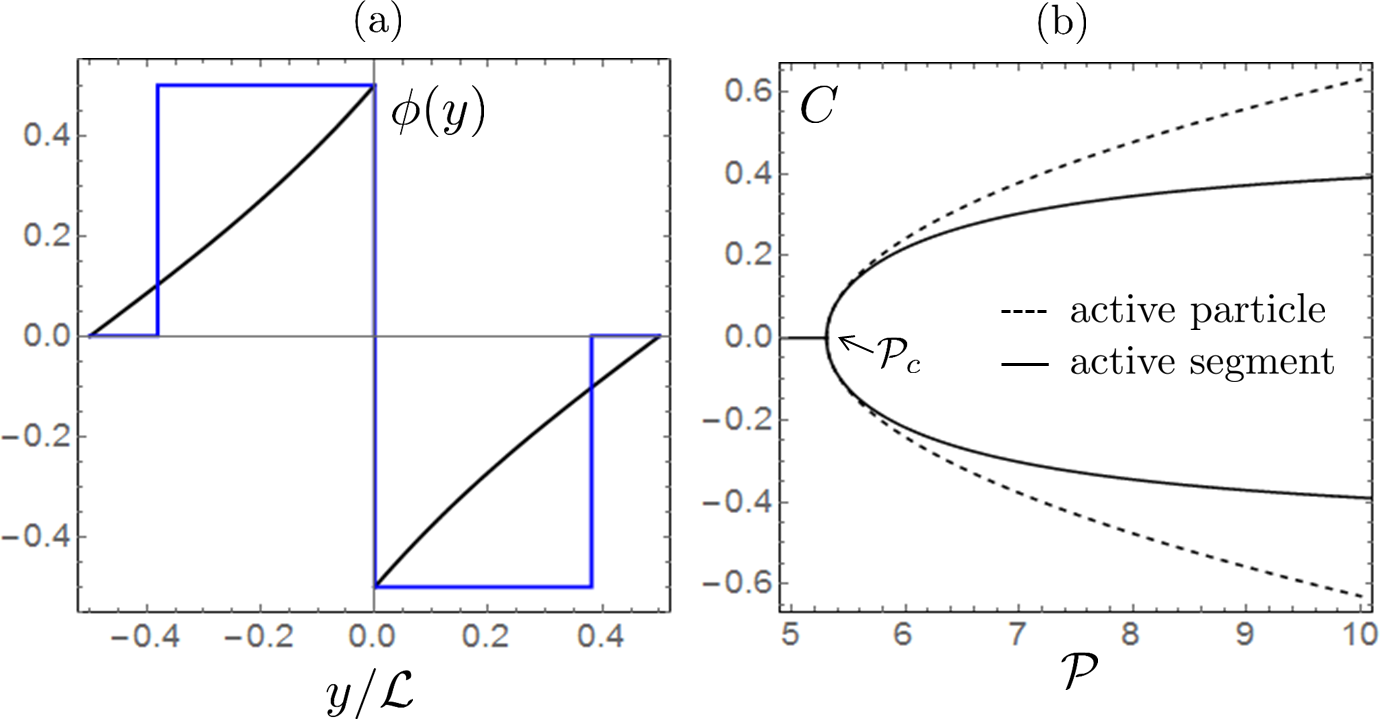}
\caption{ (a) Two AS models with different interaction  kernels $\phi(y)$: black line is the exponential kernel introduced in \cite{RecPutTru_jmps15} and used in this paper and  blue line is  the kernel used in \cite{kruse2000actively}. (b) Spontaneous polarization in  AS and AP models when $\mathcal{P}$ increases above $\mathcal{P}_c$.  Parameters: $F=0$, $\mathcal{L}=2$.}
\label{f:kernel_bifur}
\end{figure}

The flow velocity $v$ at point $y$ is induced  by  the presence in another point $z$ of an active force dipole, represented by a motor concentration-dependent active stress~\cite{RecPutTru_prl13,putelat2018mechanical}, and is also affected by a passive external force field imposed inside the cell $Ff(z,t)$ where $F$ represents the total applied force and $\int_{-\mathcal{L}/2}^{\mathcal{L}/2}f(z,t)dz=1$. If the implied nonlocal interaction, that can be, for instance, of hydrodynamic origin \cite{malgaretti2017bistability}, is linear, we can write
\begin{equation}\label{eq:model_drift1}
v= -\dot S+ 
(\mathcal{P}/\mathcal{L})\, \phi \ast c + F\, (\phi \ast \partial_yf + f),
\end{equation}
where we introduce the convolution: $\phi \ast h(y,t)= \int_{-\mathcal{L}/2}^{\mathcal{L}/2}\phi(y-z)h(z,t)dz$.  
The interaction kernel $\phi(z)$, whose detailed expression follows from finding the stress by solving the equation of local momentum balance equipped with an appropriate  constitutive law~\cite{SI}, must be an odd  function  to ensure that a symmetric distribution of force dipoles does not generate a directional  flow. In Fig.~\ref{f:kernel_bifur}(a) we illustrate two physically motivated examples of such kernels  introduced in \cite{kruse2000actively,RecPutTru_jmps15}. The dimensionless parameter $\mathcal{P}>0$ characterizes  motor contractility.   To find the unknown fronts velocity $\dot{S}$ we use the condition of impenetrability at the cell membrane $v\vert_ {\pm \mathcal{L}/2}=0$. 

Suppose now  that the internal configuration of the motors $c(y,t)$ is not observable and that we have  access only to  some global polarity measure $C(t)=\left\lbrace \phi\ast c \right\rbrace/\mathcal{L}\in[-\max \phi,\max \phi]$,  where  $\left\lbrace h\right\rbrace =(h\vert_ {\mathcal{L}/2}+h\vert_ {-\mathcal{L}/2})/2$. To obtain a closed description of the cell dynamics in terms of the `macro-variables' $C(t)$ and $S(t)$, we need to map the AS model \eqref{eq:model_motors}--\eqref{eq:model_drift1} onto an AP model.  
 
To this end, we first average \eqref{eq:model_drift1} in two different ways. Using the impenetrability condition we have $\dot{S}=\mathcal{P} C +  F\left\lbrace \phi\ast \partial_yf\right\rbrace$ and by integrating \eqref{eq:model_drift1} over space we obtain $\langle v \rangle=-\dot{S}+F/\mathcal{L}$. To eliminate the new macroscopic variable $\langle v \rangle$ we mimic \eqref{eq:model_motors} by writing $\dot{C}+\langle v \rangle=-\Phi(C)$. Here, the term $\langle v \rangle$ can be viewed as the analog of the drift term in \eqref{eq:model_motors}, in particular,  it ensures  that a retrograde flow with $\langle v \rangle<0$ contributes to the growth of polarity. The term on the right-hand side is intended to play the role of diffusion degrading the  existing polarity and therefore the  function $\Phi$ is chosen to be increasing and vanishing at $C=0$. We obtain the closed nonlinear system of ordinary differential equations:
\begin{equation}\label{eq:meta_model_231}
 \dot S=\mathcal{P} C + k_S  F,\,\ \dot {C} =\mathcal{P} C+k_C   F -\Phi(C),
\end{equation}
where $k_S =\left\lbrace \phi\ast \partial_yf\right\rbrace$ and $k_C= k_S-1/\mathcal{L}$. Note that the AP model~\eqref{eq:meta_model_231} is non-potential because  the cell position depends on its polarity while  the reverse effect is absent.  To relate the AP and AS models quantitatively we need to find a relation between the functions $\phi(y)$  and $\Phi(C)$. 

For determinacy,  we choose from now on to work with a particular AS model describing a contracting active gel on a solid background \cite{BoiJulGri_prl11, RecJoaTru_prl14} which is characterized by the kernel:  $\phi(y)= \text{sh}\left(y+\mathcal{L}/2\right) /\left[2\text{sh}(\mathcal{L}/2)\right]-H(y)\text{ch}\left(y\right)$, where $H(y)$ is the Heaviside function, see   Fig.~\ref{f:kernel_bifur}~(a) and  \cite{SI} for details. To find a matching function $\Phi(C)$, it is sufficient to consider the case $F=0$. 

Under these conditions, when the contractility parameter $\mathcal{P}$ in the AS model increases above a critical threshold $\mathcal{P}_c(\mathcal{L})$, the symmetric homogeneous solution of \eqref{eq:model_motors}--\eqref{eq:model_drift1}  $v=0$, $c\equiv 1$ and $V=0$ becomes unstable and a polarized motile state emerges as a result of pitchfork bifurcation (second order phase transition) leading to two symmetric configurations  with opposite polarities, see \cite{RecJoaTru_prl14,SI} and Fig.~\ref{f:kernel_bifur}~(b).  

To reproduce the same bifurcation in the framework of the AP model, we assume that $\Phi(C)=\partial_C\bar W$, where  $\bar W(C)=\alpha C^4/4+\mathcal{P}_cC^2/2$ is the standard expression from the theory of second order phase transitions. At $ F=0$ the polarity evolves according to the equation $\dot C=-\partial_CW$, where now $W(C)=\alpha C^4/4-(\mathcal{P}-\mathcal{P}_c)C^2/2$ is the corresponding Landau potential. It has a single minimum at $C=0$ when $\mathcal{P}<\mathcal{P}_c$ and two symmetric minima at $C=\pm \sqrt{(\mathcal{P}-\mathcal{P}_c)/\alpha}$ when $\mathcal{P}>\mathcal{P}_c$, see Fig.~\ref{f:kernel_bifur}~(b). The coefficient $\alpha$ can be fixed by matching the asymptotic behavior for the two models at $\mathcal{P}=\mathcal{P}_c$. From a normal form analysis of the AS model, we obtain $\alpha=\mathcal{P}_c^2\mathcal{L}^3\theta_2 (\mathcal{L})/2$ where the analytical expression for the function $\theta_2 (\mathcal{L})>0$ is given in~\cite{SI}.

To test the efficiency of our calibration procedure, we now subject both systems, AS and AP, to a fixed external force and compare the resulting velocity-force \mbox{(V-F)} relations for steady regimes. In the AS framework, for simplicity, we will deal with the case when where the external force  is shared by the two  boundaries of the segment: $f(y,t)=\beta \delta(y+\mathcal{L}/2)+(1-\beta)\delta(y-\mathcal{L}/2)$ where the value of $\beta$ is irrelevant in what follows because of the fixed length constraint. 
For the corresponding AP model we obtain $k_S(\mathcal{L})= (1/2)\coth(\mathcal{L}/2)>0$. 

\begin{figure}[h!]
\centering
\includegraphics[scale=0.35]{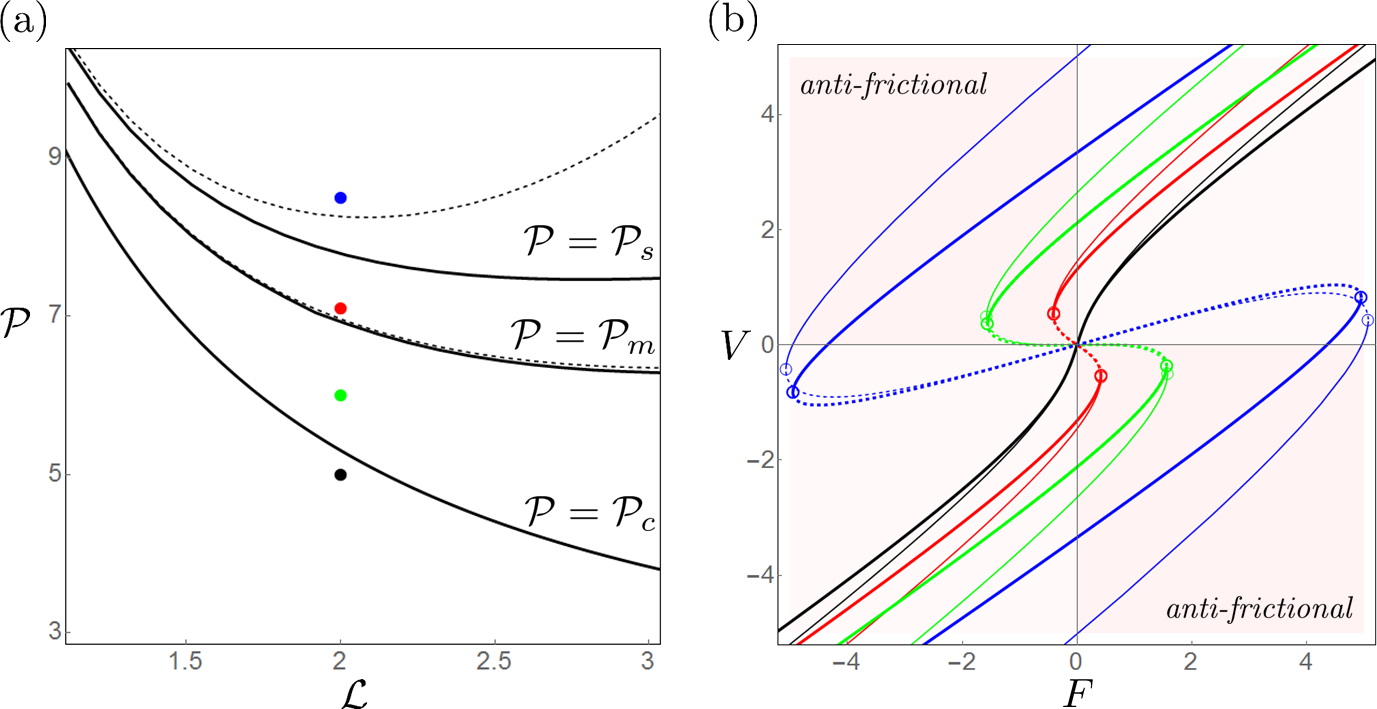}
\caption{Comparison of the V-F relations in AS and AP models. (a) The  three thresholds $\mathcal{P}_c$, $\mathcal{P}_m$  and $\mathcal{P}_s$ as   functions of the parameter $\mathcal{L}$. Thick lines -  AS model,   dotted lines - AP model. The value of $\mathcal{P}_c$ is the same in both models by construction. 
(b) Four typical V-F relations in the AS (thick lines) and the AP (thin lines) models.  The dashed parts of the V-F curves correspond to unstable regimes.  Parameters $\mathcal{L}=2$ and $\mathcal{P}=\textcolor{black}{5}$ (black),  $\mathcal{P}=6$ (red), $\mathcal{P}=7$ (green) and $\mathcal{P}=9$ (blue) are represented with color dots on panel (a). }
\label{f:velocity_force}
\end{figure}

In the case of the AS model, we solve  numerically equations \eqref{eq:model_motors}--\eqref{eq:model_drift1} with  $\partial_tc=0$. In the AP setting we find  the stationary value of polarity $C$  directly from the equation 
$\partial_CW= k_C  F$ and then obtain the V-F relation  substituting this value of $C$ into \eqref{eq:meta_model_231}.  As shown on Fig.~\ref{f:velocity_force}~(b), both models generate quantitatively similar V-F  relations in the whole range of  parameters. 

When $\mathcal{P}<\mathcal{P}_c$, the V-F relations in both models are single-valued and frictional, meaning that $V F >0$.   This is obvious  in the AP  case since  the potential $W(C)$ is convex  and the system has only one  stable ($\partial_{CC}W(C_0)>0$) stationary solution $C_0(F)$. The ensuing V-F  relation can be written explicitly  $V =k_S F+\mathcal{P} C_0( F)$. Note that due to the presence of motors, the effective  viscosity  in the active system $\mu_0=\partial_{V} F\vert_{V=0}=(\mathcal{P}_c-\mathcal{P})/(k_S\mathcal{P}_c-\mathcal{P}/\mathcal{L})$  is smaller than in its passive analog ($\mathcal{P}=0$) and can even reach zero,  which is a  feature of many active systems \cite{haines2008effective, lopez2015turning}. A similar but less explicit analysis can be performed for the AS model, see \cite{SI}. 

When $\mathcal{P}>\mathcal{P}_c$,  the V-F curves  develop a domain of bi-stability which spreads over a range $F\in[- F _t, F _t]$, where, in the AP model, $F_t=2(\mathcal{P}-\mathcal{P}_c)^{3/2}/(3k_C\sqrt{3\alpha})$.  Within this range, the stationary polarity can take three values: $C_0^* <C_0<C_0^{**} $  where $C_0^*<0<C_0^{**}$ correspond to metastable  solutions and $C_0$ is an unstable solution ($\partial_{CC}W(C_0)<0$). In this range, the \mbox{V-F} relations allow for  the coexistence of the two  metastable regimes with different signs of velocity:   $V^{*} =k_S F+\mathcal{P} C_0^{*}(F)$ and $V^{**} =k_S F+\mathcal{P} C_0^{**}(F)$. These two branches of the \mbox{V-F} relation are connected by the unstable branch $V_0=k_S F +\mathcal{P} C_0(F)$, which is located between the two turning points $F=\pm  F_{t}$. Inside  the coexistence interval $[- F_t, F_t]$, one of the two metastable solutions necessarily operates in an anti-frictional regime with $VF\leq 0$ \cite{SI}. Similar bi-directionality  with negative viscosity at zero force, is also characteristic of the V-F curves describing an  ensemble of molecular motors   interacting  either hydrodynamically  \cite{malgaretti2017bistability} or through a rigid backbone \cite{julicher1995cooperative}.

A new feature of the model is the existence of another threshold, $\mathcal{P}_m$ ($=k_S\mathcal{L}\mathcal{P}_c$ in  the AP case), beyond which the zero velocity regime stabilizes, the viscosity at zero force becomes positive again and   the V-F curves  start to display  muscle-like  stall force states. For $\mathcal{P}_m<\mathcal{P}<\mathcal{P}_s$,  where $\mathcal{P}_s=2k_S\mathcal{P}_c/(3/\mathcal{L}-k_S)$ in the AP case, such states are unstable but for $\mathcal{P}>\mathcal{P}_s(\mathcal{L})$ they stabilize.
  The functions  $\mathcal{P}_{m,s}(\mathcal{L})$ for the AS model  are compared with those for the AP model in Fig.~\ref{f:velocity_force}(a) and the corresponding  V-F curves can be read off Fig.~\ref{f:velocity_force}(b).
\begin{figure}
\centering
\includegraphics[scale=0.5]{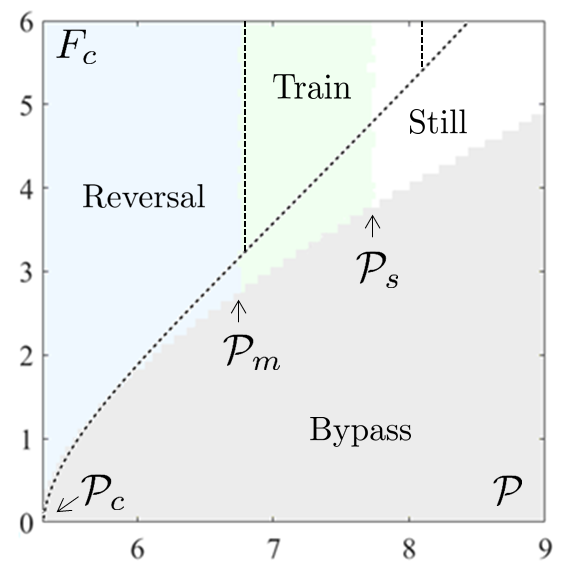}
\caption{ Phase diagram showing the four robust outcomes of the collision test: Reversal, Train and Pairing of cells which splits into a motile phase (Train) and a static one (Still). Typical dynamics are illustrated on Fig.~\ref{f:dyna_model_full}. Parameters are $\mathcal{L}=2$ and $\mathcal{D}_c=0.1$. Coloring corresponds to the AS  model  while dashed lines are  for the AP model. }
\label{f:phase_diag}
\end{figure}
 
Consider now two identical cells moving towards each other. The cells will be represented either as AS or AP  and we shall use the subscripts  $-$  and $+$  to differentiate between cells approaching  from the left and from the right. We assume that the colliding cells have to overcome  the repulsive force  \begin{equation}\label{e:force_col}
 F_{\pm}(\mathcal{D})=\pm  F_c\exp(-\mathcal{D}/\mathcal{D}_c),
\end{equation}
which depends on the separation:   $\mathcal{D}(t)=|x_{r_+}(t)-x_{f_-}(t)|$ for the AS model and $\mathcal{D}(t)=|S_+(t)-S_-(t)|$ for the AP model. Note that we introduce a characteristic  size of a cell-cell contact  $\mathcal{D}_c\ll \mathcal{L}$ and define $F_c$ as the scale of the repulsive force. We also implicitly assume  that the adhesive clusters are only transient and cannot support significant tensile loads \cite{stramer2017mechanisms}. 
\begin{figure}
\centering
\includegraphics[scale=0.4]{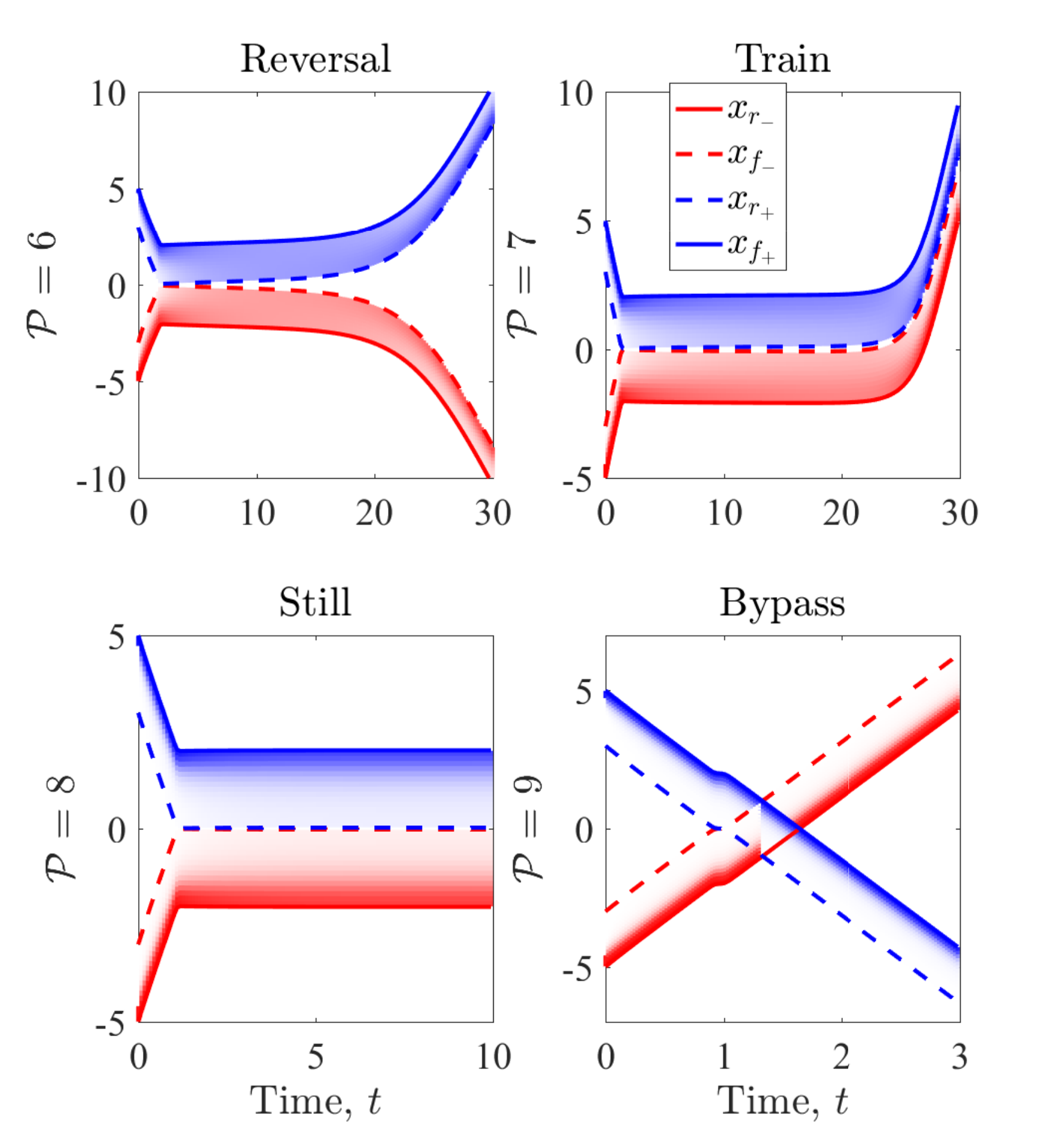}
\caption{The configurations of two colliding cells obtained within the AS model  (\ref{eq:model_motors},\ref{eq:model_drift1})  with the  contact force \eqref{e:force_col}. The intensity  of the coloring between the front lines is proportional to the  concentration of molecular motors. Parameters: $\mathcal{L}=2$, $\mathcal{D}_c=0.1$ and $F_c=4.5$. }
\label{f:dyna_model_full}
\end{figure}

In our simulations we explore the whole range  $\mathcal{P}>\mathcal{P}_c$.  Similar to what is observed in experiment \cite{desai2013contact, scarpa2013novel} and predicted based on a much more detailed model \cite{kulawiak2016modeling}, we record four possible outcomes of  collision tests: \emph{reversal},  paring, which can be motile (\emph{train}) or static (\emph{still}),   and  \emph{bypass}.  Our results are presented in the $(\mathcal{P},  F_c)$ phase diagram shown for both AS and AP models in Fig.~\ref{f:phase_diag}. 
 
In the reversal regimes, the active objects  re-polarize during collision as a result of being exposed to sufficiently large contact force.  The outcome of such `quasi-elastic' collision is that the colliding agents change the signs but not the magnitudes of their velocities. In the bypass regimes, the agents go past each other because the model allows for mutual overlap and the  contact force is not sufficient to impede the  propulsive machinery. In the pairing  regimes the two initially mobile agents first get immobilized and push against each other as both of them reach  transiently stall conditions. Such regimes can be stable (forming a robust still phase) only for $\mathcal{P}>\mathcal{P}_s$ where a steady stall state exists. For $\mathcal{P}<\mathcal{P}_s$, in the train regimes, one of the two active agents eventually sweeps along the other by repolarizing its internal configuration and afterwards they continue traveling together as a cell doublet (train). In these regimes, stall configurations do exist but are unstable and are destroyed by infinitesimal perturbations, see \cite{SI} for details. Note however that the exact partitioning of the pairing phase into still and train regimes is somewhat arbitrary  since we did not specify the attractive/cohesive structure of the interaction model. 

\begin{center}
\begin{figure}[h]
\includegraphics[scale=0.40]{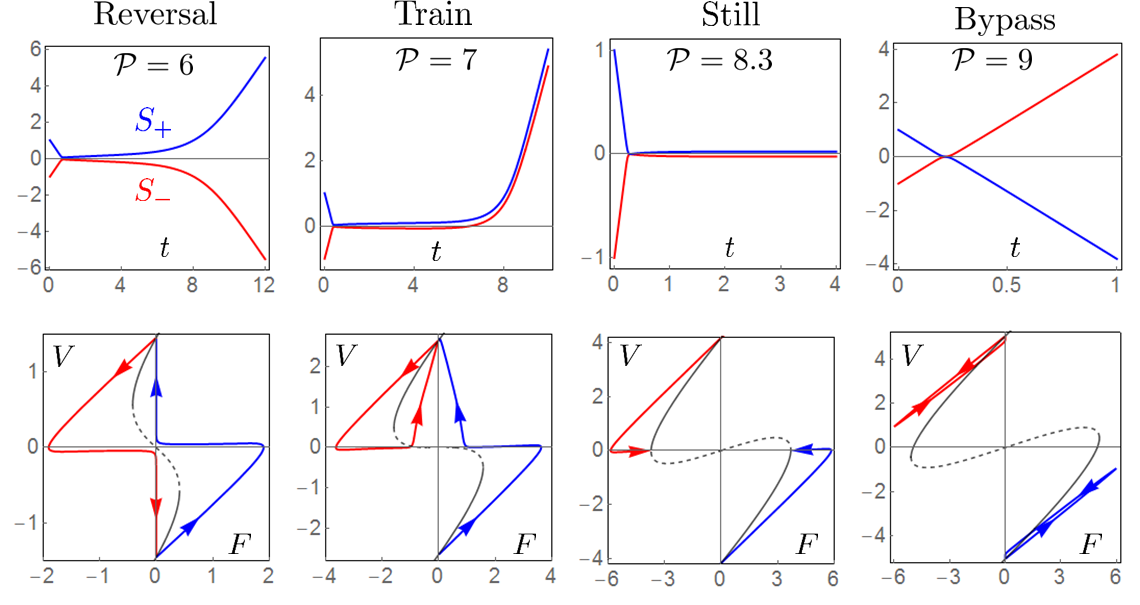}
\caption{Four typical collision scenarios observed in the AP model. The first row shows the trajectories of the colliding  particles. The second row illustrates the transient velocity-force dynamics superimposed on the steady V-F relation. Parameters: $\mathcal{L}=2$, $F_c=6$ and $\mathcal{D}_c=0.1$}
\label{f:active_particles_dyna}
\end{figure}
\end{center}

The typical trajectories of the colliding active agents for each of the four regimes  and the corresponding  configurations of molecular motors in the AS and AP models are illustrated in Fig.~\ref{f:dyna_model_full} and Fig.~\ref{f:active_particles_dyna}. Remarkably, both models show a qualitatively similar behavior. In the AP case,  we also represent the transient collision trajectories $( F_{\pm}(t),\dot{S}_{\pm}(t))$ on the V-F plane where they can be compared with the stationary V-F relations. The two types of curves deviate because, in collision tests, polarity does not relax instantaneously to its steady value; moreover the interaction force itself varies  during such relaxation. The finite time of relaxation is particularly important for the existence of  train regimes  which  crucially rely on the ability of active agents to  transiently reach an unstable stall state.  

To conclude,  we have used a one-dimensional model of contraction-driven cell  motility to develop an equivalent active particle model spontaneously adjusting its polarity to the applied force. Both models are able to predict the outcomes of collision tests covering the whole spectrum of observed behaviors. An important prediction is that cell contractility serves as the key internal regulator of CIL which could be further investigated experimentally by using drug treatments \cite{mitrossilis2009single} or optogenetics \cite{valon2017optogenetic}. The proposed model of active particle with self-adjusting polarity is capable of describing complex mechanical cell-cell interactions and can prove useful for the development of a kinetic theory of tissues driven by internal cellular motion. 

\acknowledgments{P.R. acknowledges support from a CNRS-Momentum grant. T.P. was supported by the EPSRC Engineering Nonlinearity project No. EP/K003836/1. L.T. is grateful to the French government which supported his work under Grant No. ANR-10-IDEX-0001-02 PSL.}




%

\end{document}